\documentclass[aps,prd,tightenlines,twocolumn,superscriptaddress]{revtex4-1}

\usepackage{amsmath,amssymb,amsfonts,bm}
\usepackage{graphicx}
\usepackage{dcolumn}
\usepackage[colorlinks=true,linkcolor=blue]{hyperref}

\begin{document}


\title{Neutrino oscillation from the beam with Gaussian-like energy distribution }

\author{Rong-Sheng Han}
\email{hrs@ncepu.edu.cn}
\affiliation{Mathematics and Physics Department, North China Electric Power University, Beijing, 102206, China}
\author{Liang Chen}
\affiliation{Mathematics and Physics Department, North China Electric Power University, Beijing, 102206, China}
\author{Ke-Lin Wang}
\affiliation{Department of Modern Physics, University of Science and Technology of China, Hefei, 230026, China}

\date{\today}

\begin{abstract}
 A recent neutrino experiment at Daya Bay gives superior data of the distribution of the prompt energy. In this paper, the energy distribution presented in the experiment is simulated by applying a Gaussian-like packet to the neutrino wave function received by the detector. We find that the wave packet of neutrinos is expanded during the propagation. As a result, the mixing angle $\theta_{13}$ is more difficult to be measured than $\theta_{12}$ and $\theta_{23}$ in long baseline experiments.
 Some other propagation properties, such as the time evaluation of the survival probability, the neutrino oscillation and the $CP$ violation, are also studied with the employment of the coherent state method. When the Gaussian packet width increases, the amplitude of the neutrino oscillation decreases, whereas the oscillation period increases gradually. 
\end{abstract}


\maketitle


\section{Introduction}\label{section1}

Neutrino masses and flavor mixing have attracted much attention in the past few years in fundamental physics\cite{ChinPhysC.38.090001,PhysRevD.86.010001}. The mixture is governed by the $3\times3$ unitary Pontecorvo-Maki-Nakagawa-Sakata (PMNS) matrix\cite{Pontecorvo1,Pontecorvo2, Maki01111962} with three mixing angles $(\theta_{12}, \theta_{13}, \theta_{23})$ and a $CP$-violating phase $\delta$. Theoretically, the $CP$ violation and the neutrino oscillation are experimental observable if the mixing angles $\theta_{ij}$ and the squared mass differences $\Delta{m}_{ij}^2=m_{i}^2-m_{j}^2$ are nonzero\cite{ChinPhysC.38.090001}.

Very recently, the nonzero mixing angle $\theta_{ij}$ has been established for $\theta_{12}$ and $\theta_{23}$\cite{PhysRevD.86.010001,gonzalez2014nufit,JHighEnergyPhys.2012.123} with high accuracies. Relatively, $\theta_{13}$ is more difficult to be determined in experiments than other two angles. G. L. Fogli {\it et al.} found the strong evidence ( $\sim 3\sigma$ level, where $\sigma$ is the standard deviation) for the nonzero $\theta_{13}$ in 2011\cite{PhysRevD.84.053007}. Then the short-baseline reactor experiments in Daya Bay\cite{PhysRevLett.108.171803} and in RENO\cite{PhysRevLett.108.191802} improved the confidence level up to $5\sigma$ for $\theta_{13}>0$. The Daya Bay experiment found that $\sin^2\theta_{13}\approx0.023{\pm}0.003$  via the survival probability of the electron antineutrino $\bar{\nu}_{e}$ at short distances from the reactor, which is given as $P_{\text{sur}}=1-\sin^2(2\theta_{13})\sin^2(1.267\Delta{m}_{13}^2\cdot{L/E})$, where $E$ is the neutrino energy and $L$ is the travel distance. The RENO experiment received a very close result $\sin^2\theta_{13}\approx0.029{\pm}0.006$ via a similar setup. However, the thoretical investigation for short distance measurement is still incomplete. In this paper, we use the wave packet method to study the neutrino oscillation, and find that (1) the oscillation is dominated by $\theta_{13}$ in short distance, (2) the wave packet decay exponentially for large distance. Both these results demonstrate that short distance meansurement is important. 

The accurate determination of the three mixing angles is of essential importance to research the $CP$ violation in the neutrino sector. Lots of attempts are tried to improve the precision of $\theta_{13}$. The accuracy of mixing angles from oscillation measurement hangs on the availability of a more accurate expression of survival probability, of which the implementation needs fewer assumptions, thus allowing a flexible experimental arrangement, e.g., the source-detector distance is no more a critical prerequisite. Such an expression for survival probability should be obtained in a strict sense.

In both the experiments of Daya Bay and RENO, we noticed that the surviving probabilities have been given with regard to a single energy value of the neutrino, however the true neutrino beam may have a wide energy distribution. For example, the energy spectrum of the electron antineutrinos from the reactor at Daya Bay peaks at $\sim3.0$ MeV with a half width of comparable value\cite{PhysRevLett.108.171803,PhysRevLett.108.191802}.
More details about the mentioned experiments show that the neutrino beams utilized have Gaussian-like energy distributions (See figure 5 in Ref. \cite{PhysRevLett.108.171803}). Therefore, a more precise and practical method to study the neutrino properties presented in the experiments is using the Gaussian-like packet to simulate the wave function received by the detector. The wave packet approach\cite{PhysRevD.70.053010} to the neutrino oscillation has been proposed with different motivations\cite{Nauenberg199923,PhysRevD.24.110,Fuju2006,Kayser2010}, e.g., to cope with the uncertainty principle, but it has not got the deserved attention. In this work, we use the Gaussian-like packet to simulate the energy distribution of the electron neutrino of Daya Bay and RENO, and to study the neutrino oscillation properties. For a given single energy value (say $3.0$ Mev) and other reliable parameters, we find the electron neutrino oscillation period is about $\lambda_{21}=97.63$km, while the fast oscillation period is approximately $\lambda_{31}\approx\lambda_{32}=3.194$km. The $CP$ violation in this case has a same oscillation period as $\lambda_{21}$. When the energy distribution is taken into account, the survival probability of the electron neutrino which oscillates along the propagation direction is calculated, with which we demonstrate that the mixing angle $\theta_{13}$ is more difficult to be measured than $\theta_{12}$ and $\theta_{23}$ in long-baseline experiments. During the propagation, the wave packet of neutrinos gradually disperses with its fluctuation amplitude decreases and the periods simultaneously become larger. On the other hand, the similar law is found for the oscillation when in terms of the neutrino packet width. As initial packet width increases, the oscillation periods for both the survival probability and the $CP$ violation increase, whereas the amplitudes tend to be smaller.
Numerical results concerning the wave packet spreading imply that a sharp neutrino wave packet will benefit the detection of the oscillation.

The present work is organized as follows. In Sec. {\ref{section2}}, we introduce the model Hamiltonian of the one-dimensional Dirac equation for massive fermions and the parity operator, then solve the model Hamiltonian and get the propagation of the Gaussian-like packet wave function. In Sec. {\ref{section3}}, we calculate some important physical quantities: the time evaluation of three-flavor neutrino oscillations, the ratio of the density of the electron neutrino, and the oscillation of the $CP$ violation of the muon neutrino. A summary is given in Sec. {\ref{section4}}.

\section{Model Hamiltonian and wave function}\label{section2}

In the following context, we assume that the neutrino can be described by a wave packet, and the wave packet is propagating along the $z$-diretion. The momentum components orthogonal to the propagation direction are much smaller than the parallel component ($\langle{\hat{p}_x}\rangle, \langle{\hat{p}_y}\rangle \ll \langle{\hat{p}_z}\rangle$), so that we can simplify the Dirac equation $(\bm{\alpha}\cdot\hat{\bm{p}}+m\beta){\psi}=E\psi$ into a one dimensional problem:
\begin{equation} \label{eq1}
	({\alpha_z}{\hat{p}_z}+m\beta)\psi=E\psi,
\end{equation}
where the Dirac matrices take the following forms:
\begin{equation} \label{eq2}
\alpha_{j}=\left(\begin{array}{cc}
	0 & \sigma_j \\
	\sigma_j & 0 \end{array}\right) , ~~~~~~~
\beta=\left(\begin{array}{cc}
	1 & 0 \\
	0 & -1 \end{array}\right),
\end{equation}
$j=x, y, z$, $m$ is the mass of neutrino, $E$ is the eigen-energy and $\psi$ is the wave function with four components. One can find that the spin-up and spin-down states are decoupled in the one dimensional Dirac equation (\ref{eq1}). Considering the neutrino oscillation problem concentrated in this paper is irrelated to the spin degree of freedom and without lose of generality, we only take into account the spin-up states and then the Dirac equation can be simplified to be
\begin{equation} \label{eq3}
\left \{
\begin{array}{c}
	\hat{p}_z\psi_2 + m\psi_1=E\psi_1 \\
	\hat{p}_z\psi_1 - m\psi_2=E\psi_2
\end{array},
\right.
\end{equation}
where $\psi_1$, $\psi_2$ are two components of spin-up states. The conjugate operators $\hat{p}_z$ and $\hat{z}$ can be represented by the following bosonic creation and annihilation operators:
\begin{equation} \label{eq4}
\left \{
\begin{array}{rl}
	\hat{z}&=\frac{i\Delta}{\sqrt{2}}\left(b-b^{\dagger}\right) \\
    \hat{p}_z&=\frac{1}{\sqrt{2}\Delta}\left(b+b^{\dagger}\right)
\end{array},
\right.
\end{equation}
where $\Delta$ is a arbitrary number. Consequently, the model Hamiltonian (\ref{eq3}) could be rewritten as
\begin{equation}  \label{eq5}
	\hat{H}=\frac{\sigma_x}{\sqrt{2}\Delta}\left(b+b^{\dagger}\right)+m\sigma_z .
\end{equation}

Now we investigate the eigenstates of the model Hamiltonian. It is easy to check that $\hat{H}$ commutates with the parity operator $\hat{\Pi}=\exp{(i\pi\hat{N}})$, where $\hat{N}=\sigma_z+\frac{1}{2}+b^{\dagger}b$. With some tedious but straightforward derivations, it could be proven that the common eigenstates of the Hamiltonian and the parity operator can be written as
\begin{align}
|\Pi=+1\rangle=\left[\begin{array}{c}
	F_{+}\left(e^{-b^{\dagger}b^{\dagger}/2+\alpha{b}^{\dagger}}-e^{-b^{\dagger}b^{\dagger}/2-\alpha{b}^{\dagger}}\right)|0\rangle \\
	G_{+}\left(e^{-b^{\dagger}b^{\dagger}/2+\alpha{b}^{\dagger}}+e^{-b^{\dagger}b^{\dagger}/2-\alpha{b}^{\dagger}}\right)|0\rangle
	\end{array} \right]~\label{eq6}\\
|\Pi=-1\rangle=\left[\begin{array}{c}
	F_{-}\left(e^{-b^{\dagger}b^{\dagger}/2+\alpha{b}^{\dagger}}+e^{-b^{\dagger}b^{\dagger}/2-\alpha{b}^{\dagger}}\right)|0\rangle \\
	G_{-}\left(e^{-b^{\dagger}b^{\dagger}/2+\alpha{b}^{\dagger}}-e^{-b^{\dagger}b^{\dagger}/2-\alpha{b}^{\dagger}}\right)|0\rangle
	\end{array} \right]~\label{eq7} ,
\end{align}
where $F_{+}$, $G_{+}$, $F_{-}$ and $G_{-}$ are arbitrary coefficients, $\alpha$ is the eigenvalue of the annihilation operator, and $|0\rangle$ is the ground state of the operator $b^{\dagger}b$ with the formula $b^{\dagger}b|0\rangle=0$. For a given $\alpha$, energy eigenvalues could be found by solving the eigenvalue equation $\hat{H}|\Pi=\pm1,\alpha\rangle=E^{\Pi=\pm1}_{(\alpha)}|\Pi=\pm1,\alpha\rangle$, obtaining
\begin{equation}\label{eq8}
	E^{\Pi=\pm1}_{(\alpha)}=\pm\sqrt{m^2+A^2\alpha^2},
\end{equation}
where $A=1/\sqrt{2}\Delta$. The coefficients (unnormalized) corresponding to the positive and the negative eigenstate energies are
\begin{equation}\label{eq9}
F_{+(-)}=1, ~~ G_{+(-)}=\alpha/\left[\sqrt{2}\Delta(m+|E_{\alpha}|)\right], ~~ E_{(\alpha)}>0 ,
\end{equation}
\begin{equation}\label{eq10}
F_{+(-)}=-\alpha/\left[\sqrt{2}\Delta(m+|E_{\alpha}|)\right], ~~ G_{+(-)}=1, ~~ E_{(\alpha)}<0 ,
\end{equation}
where $+(-)$ denotes the parity. From Eqs. (\ref{eq9}) and (\ref{eq10}), we can see that the obtained eigenstates are two-fold degenerate due to the parity symmetry.

\begin{figure}[tb]
  \centering
  \includegraphics[width=0.8\columnwidth]{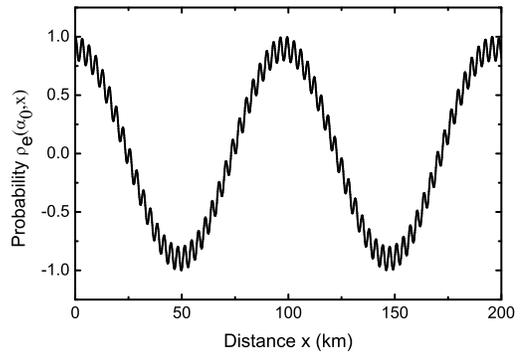}\\
  \caption{The electron neutrino oscillation as a function of the distance away from the source. The energy of the initial electron neutrino is chosen to be the center of the Gaussian packet, $A\alpha_0=3.0$MeV. } \label{fig1}
\end{figure}

It could be tested that the eigenstates Eqs. (\ref{eq6}) and (\ref{eq7}) with a single vale of $\alpha$ cannot be normalized, so that the real wave function which is physically significant should be a wave packet consisting of eigenstates with different values of $\alpha$. Considering that the neutrino beam in the experiments have a Gaussian-like energy distribution, the neutrino wave functions with definite parities are supposed to have the following forms
\begin{eqnarray}\label{eq11}
|\nu^{(\pm)}\rangle=\int{d}\alpha{e^{-\gamma(\alpha-\alpha_0)^2}|\Pi=\pm1\rangle} ,
\end{eqnarray}
where $\alpha_0$ is the center of the Gaussian distribution, $\gamma$ is determined by the half-width, $|\Pi=\pm1\rangle$ are given in Eqs. (\ref{eq6}) and (\ref{eq7}), and $F_{\pm}$ and $G_{\pm}$ are chosen to be the coefficients of the positive energy solutions in Eq. (\ref{eq9}) for simplicity.
The wave function constructed above is a stationary one which is impossible to propagate along $z$ direction.
It could be computed that the average propagation velocity of a definite parity Gaussian-like packet is exactly zero. For example, the wave function of the parity-positive state at time $t$ can be written as
\begin{equation}\label{eq12}
|\nu^{(+)}(t)\rangle=\int{d\alpha}e^{-\gamma(\alpha-\alpha_0)^2}e^{-iE_{(\alpha)}t}|\Pi=+1,\alpha\rangle .
\end{equation}
Then the expectation values of the momentum operator $\hat{p}_z$ and the position operator $\hat{z}$ are both zero:
\begin{equation}\label{eq13}
\langle\nu^{(+)}(t)|\hat{p}_z|\nu^{(+)}(t)\rangle=\langle\nu^{(+)}(t)|\hat{z}|\nu^{(+)}(t)\rangle=0 .
\end{equation}
So that the propagation wave packet must be a mixture state of the parity-positive and the parity-negative states. Without lose of generality, we choose the following initial state whose propagation velocity is maximised:
\begin{align}
&|\nu\rangle=\frac{1}{2}\left(|\nu^{(+)}\rangle+|\nu^{(-)}\rangle\right)\nonumber\\
&=\int{d}\alpha{~}{e}^{-\frac{1}{2}b^{\dagger}b^{\dagger}+\alpha{b}^{\dagger}-\gamma(\alpha-\alpha_0)^2}
	\left(\begin{array}{c}
		|0\rangle \\
		\frac{\alpha}{\sqrt{2}\Delta(m+|E_{(\alpha)}|)}|0\rangle
	\end{array} \right) . \label{eq14}
\end{align}

\begin{figure}[tb]
  \centering
  \includegraphics[width=0.8\columnwidth]{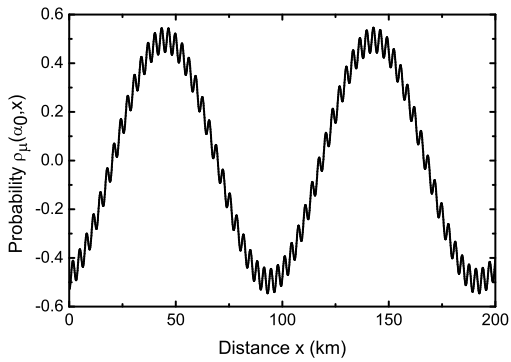}\\
  \caption{The muon neutrino oscillation as a function of distance away from the source. The energy of the initial electron neutrino is chosen to be the center of the Gaussian packet, $A\alpha_0=3.0$MeV. } \label{fig2}
\end{figure}

\begin{figure}[tb]
  \centering
  \includegraphics[width=0.8\columnwidth]{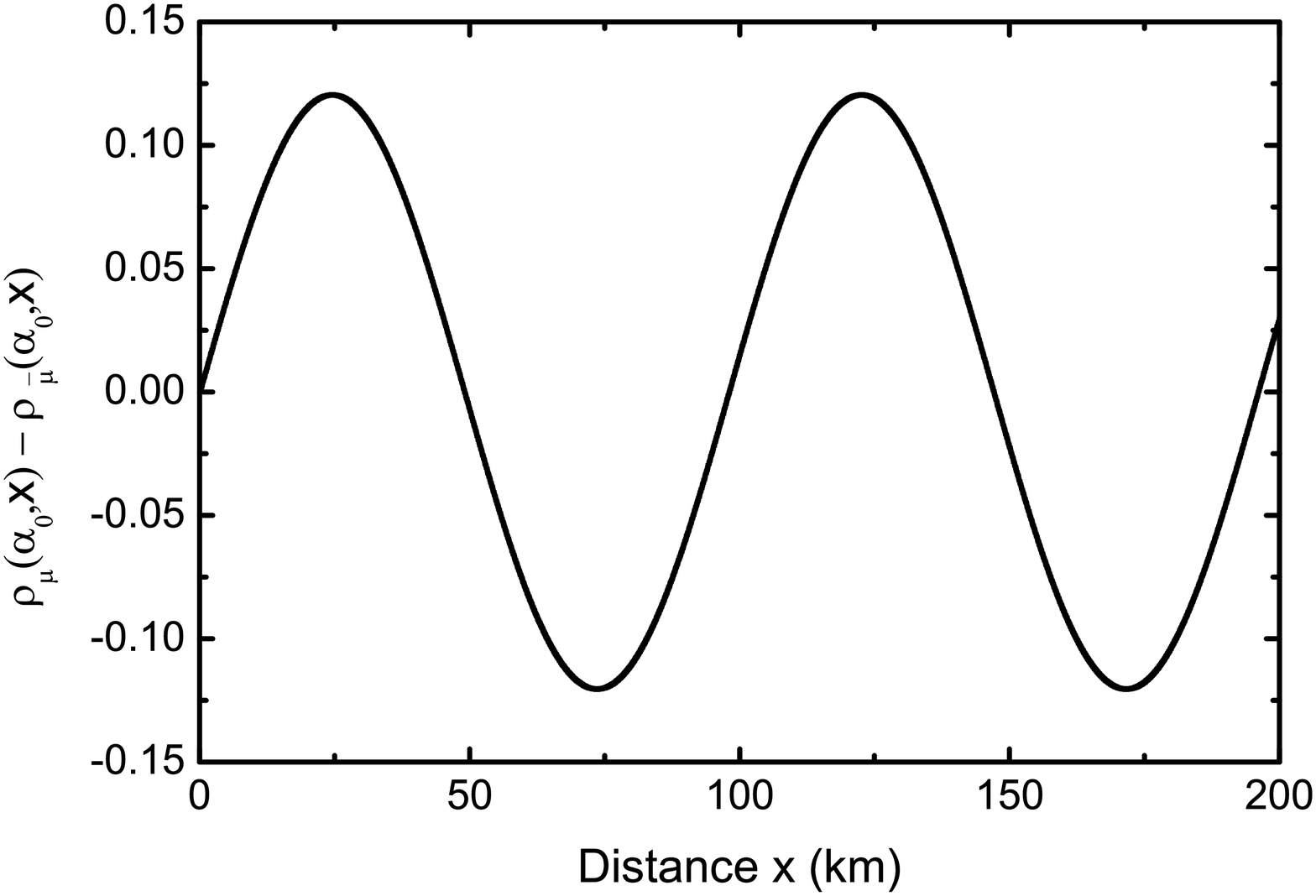}\\
  \caption{The oscillation of the $CP$ violation of the muon neutrino as a function of distance. The energy of the initial electron neutrino is been chosen to be the center of the Gaussian packet, $A\alpha_0=3.0$MeV. } \label{fig3}
\end{figure}

In experiments, the neutrinos are found to be of three flavours, $\nu_{\rho}=\left(\nu_{e}, \nu_{\mu}, \nu_{\tau}\right)$, when taking part in the weak interaction processes. Within the precision of the current experimental data, three flavour eigenstates are orthogonal with each other. Theoretically, the flavour states are linear combinations of the Dirac mass eigenstates $\nu_{j}=\left(\nu_{1}, \nu_{2}, \nu_{3}\right)$, having masses $m_{j}=\left(m_1, m_2, m_3\right)$:
\begin{eqnarray}
&\nu_{\rho}=U_{\text{PMNS}}^{\rho{j}}\nu_{j}\label{eq15} ,
\end{eqnarray}
where $U_{\text{PMNS}}$ is the PMNS lepton flavour mixing matrix
\begin{align}
&U_{\text{PMNS}}=\left(\begin{array}{ccc}
1&0&0\\
0&c_{23}&s_{23}e^{i\phi_{23}}\\
0&-s_{23}e^{-i\phi_{23}}&c_{23}
\end{array}\right)\cdot\nonumber\\
&\left(\begin{array}{ccc}
c_{13}&0&s_{13}e^{i\phi_{13}}\\
0&1&0\\
-s_{13}e^{-i\phi_{13}}&0&c_{13}
\end{array}\right)\cdot
\left(\begin{array}{ccc}
c_{12}&s_{12}e^{i\phi_{12}}&0\\
-s_{12}e^{-i\phi_{12}}&c_{12}&0\\
0&0&1
\end{array}\right) , \label{eq16}
\end{align}
in which $\rho=\left(e, \mu, \tau\right)$,  $c_{ij}=\cos{\theta_{ij}}$, $s_{ij}=\sin{\theta_{ij}}$, $i,j=1,2,3$, $\theta_{ij}$ are the mixing angles, and $\phi_{ij}$ are the $CP$ violation angles. Finally, our initial state of $\nu_e$ neutrino can be rewritten as
\begin{align}\label{eq17}
&|\nu_e(t=0)\rangle=N\int{d\alpha}e^{-\gamma(\alpha-\alpha_0)^2}\sum_{j=1,2,3}U_{\text{PMNS}}^{ej}|\nu_{j(\alpha)}\rangle ,\nonumber\\
&|\nu_{j(\alpha)}\rangle=e^{-\frac{1}{2}b_j^{\dagger}b_j^{\dagger}+\alpha{b}_j^{\dagger}}\left(\begin{array}{c}
|0\rangle \\
\frac{\alpha}{\sqrt{2}\Delta(m_j+E_{\alpha}^j)}|0\rangle
\end{array}\right),
\end{align}
where $N$ is the normalization constant. In the ultrarelativistic limit, the masses of neutrinos are much smaller than their energies $m_j\ll{E_{j(\alpha)}}\sim A\alpha$, so we can take the following approximation,
\begin{align}
&E_{j(\alpha)}=A\alpha+\frac{1}{2}\frac{m_j^2}{A\alpha}\label{eq18}\\
&\frac{\alpha}{\sqrt{2}{\Delta}(m_j+E_{j(\alpha)})}=\frac{1}{1+m_j/A\alpha}\label{eq19} .
\end{align}
Accordingly, the normalisation constant and the time evaluation of the wave function can be described as,
\begin{align}
&\frac{1}{N^2}=\pi\sqrt{\frac{4}{4\gamma-1}}{e^{2\gamma\alpha_0^2/(4\gamma-1)}}\left[1+\frac{c_{12}^2c_{13}^2}{(1+m_{1}/A\alpha_0)^2}\right.\nonumber\\
&\left.+\frac{c_{13}^2s_{12}^2}{(1+m_{2}/A\alpha_0)^2}+\frac{s_{13}^2}{(1+m_{3}/A\alpha_0)^2}\right] ,\label{eq20}\\
&|\nu_e(t)\rangle=N\int{d\alpha}e^{-\gamma(\alpha-\alpha_0)^2}\left[c_{12}c_{13}e^{-iE_{1(\alpha)}t}|\nu_{1(\alpha)}\rangle\right. \nonumber\\
&+\left.c_{13}s_{12}e^{i\phi_{12}}e^{-iE_{2(\alpha)}t}|\nu_{2(\alpha)}\rangle+s_{13}e^{i\phi_{13}}e^{-iE_{3(\alpha)}t}|\nu_{3(\alpha)}\rangle\right] .\label{eq21}
\end{align}

\section{Physical quantities}\label{section3}

Based on the time evaluation of the wave function Eq. (\ref{eq21}), we study some interesting physical quantities. The parameters used in this paper are consistent with current measurements: (1) the mass differences $\Delta{m_{12}^2}=7.59\times10^{-5}\text{eV}^2$\cite{PhysRevL.101.111301}, $\Delta{m_{32}^2}=2.32\times10^{-3}\text{eV}^2$\cite{PhysRevL.106.181801}, $\Delta{m_{31}^2}=\Delta{m_{32}^2}$; (2) mixing angles $\sin^2{2\theta_{12}}=0.861$\cite{PhysRevL.107.041801}, $\sin^2{2\theta_{13}}=0.092$\cite{PhysRevLett.108.171803}, $\sin^2{2\theta_{23}}=1.0$\cite{PhysRevL.107.041801}; (3) $CP$ phases $\phi_{12}=\phi_{23}=0$, $\phi_{13}=\pi/3$\cite{EurophysL.101.111301}; (4) Parameters that related to the Gaussian packet are estimated from the results given in Ref. \cite{PhysRevLett.108.171803} as $A\alpha_0=3.0\text{MeV}$, $\gamma=0.056\text{MeV}^2$.

\begin{figure}[tb]
  \centering
      \includegraphics[width=0.9\columnwidth]{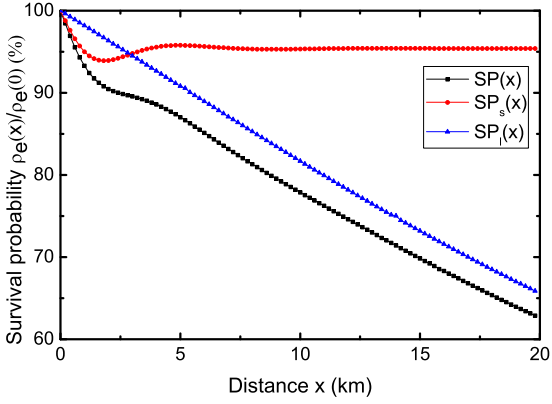}\\
  \caption{The survival probability of the electron neutrino $SP(x)=\rho_{e}(x)/\rho_e(0)$ as a function of distance for the general case (the blue line), the short range approximation $\theta_{12}=\theta_{23}=0$ (the pink line), and the long range approximation $\theta_{13}=0$ (the yellow line). A Gaussian-like wave function is applied to simulate the neutrino beam.} \label{fig4}
\end{figure}

Firstly, we consider the time evaluation of the probability of an initial electron neutrino $\rho_e(\alpha,t)=|\psi_e(\alpha,t)|^2$, where the amplitude $\psi_e(\alpha,t)=\langle\nu_e|\nu(t)\rangle$. After some tedious but straightforward derivations on the basis of the Eq. (\ref{eq17}), we found
\begin{align}
&\rho_e(\alpha,t)=N^2e^{-2\gamma(\alpha-\alpha_0)^2}\left[c_{12}^4c_{13}^4+c_{13}^4s_{12}^4+s_{13}^4\right.\nonumber\\
&\left.+2c_{13}^4c_{12}^2s_{12}^2\cos\left(\Omega_{12}t\right)+2c_{12}^2c_{13}^2s_{13}^2\cos\left(\Omega_{13}t\right)\right.\nonumber\\
&\left.+2s_{12}^2c_{13}^2s_{13}^2\cos\left(\Omega_{23}t\right)\right] ,\label{eq22}
\end{align}
and then the total probability corresponding to the Gaussian-like distributed electron neutrino beam reads
\begin{align}
&\rho_{e}(t)=\int{\rho_{e}(\alpha,t)\mathtt{d}\alpha}=C_1 \nonumber\\
&+2{N^2}\int{\mathtt{d}\alpha}e^{-2\gamma(\alpha-\alpha_0)^2}
\left[c_{13}^4c_{12}^2s_{12}^2\cos\left(\Omega_{12}t\right)\right.\nonumber\\
&\left.+c_{12}^2c_{13}^2s_{13}^2\cos\left(\Omega_{13}t\right)+s_{12}^2c_{13}^2s_{13}^2\cos\left(\Omega_{23}t\right)\right]\label{eq23} ,
\end{align}
where $C_1$ is a time (distance)-independent constant, which is given by
\begin{equation}
C_1 = N^2\sqrt{\frac{\pi}{\gamma}}\left(c_{12}^4c_{13}^4+c_{13}^4s_{12}^4+s_{13}^4\right)\label{eq24} ,
\end{equation}
and the three oscillation frequencies are defined as $\Omega_{ij}=\left(m_i^2-m_j^2\right)/{(2A\alpha)}$, ($i,j=1,2,3$).

In practical calculations, the parameter $\alpha$ has the dimension eV/$A$, where $m_{j}\ll{A\alpha}$. We can use $A\alpha$ as the leading order approximation of the energy. In addition, the speed of neutrino propagation is chosen to be the speed of light. Fig. (\ref{fig1}) gives the density evaluation of the electron neutrino with the given energy $A\alpha_0=3.0$MeV (the center of the Gaussian-like energy distribution in the Daya Bay experiment) as a function of distance. In the plot, the slow oscillation period reflects the mismatch of the masses difference $\Delta m_{12}^2$, while the fast oscillation is governed by $\Delta m_{13}^2$ and $\Delta m_{23}^2$. More carefully, the oscillation period is estimated by $\lambda_{ij}=4\pi{A}\alpha_0/\Delta{m}_{ij}^2$, which gives $\lambda_{21}=97.63$km and $\lambda_{31}\approx\lambda_{32}=3.194$km.

Similarly, we can also calculate the oscillation of the other two flavor neutrinos. We find that the corresponding probabilities of the muon neutrino and tauon neutrino are given by
\begin{widetext}
\begin{align}
\label{eq25} \rho_{\mu}(\alpha,t)=&C_2+2N^2{e}^{-2\gamma(\alpha-\alpha_0)^2}\left[\right.
c_{12}^2c_{13}^2s_{12}^2(s_{13}^2s_{23}^2-c_{23}^2)\cos(\Omega_{12}t)-c_{12}^2c_{13}^2s_{13}^2s_{23}^2\cos(\Omega_{13}t)-c_{13}^2s_{12}^2s_{13}^2s_{23}^2\cos(\Omega_{23}t)\nonumber\\
&+c_{12}c_{23}s_{13}s_{23}c_{13}^2s_{12}^2\cos(\Omega_{12}t-\phi_{12}-\phi_{23}+\phi_{13})-c_{23}s_{12}s_{13}s_{23}c_{12}^2s_{13}^2\cos(\Omega_{12}t+\phi_{12}+\phi_{23}-\phi_{13})  \nonumber\\
&-c_{12}c_{23}s_{12}s_{13}s_{23}c_{13}^2\cos(\Omega_{13}t-\phi_{12}-\phi_{23}+\phi_{13})+c_{12}c_{23}s_{12}s_{13}s_{23}c_{13}^2\cos(\Omega_{23}t-\phi_{12}-\phi_{23}+\phi_{13}) \left.\right] , \\
\label{eq26}\rho_{\tau}(\alpha,t)=&C_3+2N^2{e}^{-2\gamma(\alpha-\alpha_0)^2}\left\{c_{12}^2c_{13}^2s_{12}^2(s_{13}^2c_{23}^2-s_{23}^2)\cos(\Omega_{12}t)-c_{13}^2c_{23}^2s_{13}^2[c_{12}^2\cos(\Omega_{13}t)+s_{12}^2\cos(\Omega_{23}t)]\right.\nonumber \\
&+c_{12}^3c_{13}^2c_{23}s_{12}s_{13}s_{23}\cos(\Omega_{12}t+\phi_{12}+\phi_{23}-\phi_{13})
-c_{12}c_{13}^2c_{23}s_{12}^3s_{13}s_{23}\cos(\Omega_{12}t-\phi_{12}-\phi_{23}+\phi_{13})\nonumber \\
&+c_{12}c_{13}^2c_{23}s_{12}s_{13}s_{23}\cos(\Omega_{13}t-\phi_{12}-\phi_{23}+\phi_{13})
\left.-c_{12}s_{13}^2c_{23}s_{12}s_{13}s_{23}\cos(\Omega_{23}t-\phi_{12}-\phi_{23}+\phi_{13})\right\} ,
\end{align}
where $C_2$ and $C_3$ are time(distance)-independent constants
\begin{align}
\label{eq27}C_2=&N^2{e}^{-2\gamma(\alpha-\alpha_0)^2} \left[2c_{12}^2c_{13}^2c_{23}^2s_{12}^2 + c_{12}^4c_{13}^2s_{13}^2s_{23}^2+c_{13}^2s_{12}^4s_{13}^2s_{23}^2 + c_{13}^2s_{13}^2s_{13}^2s_{23}^2 \right. \nonumber\\
&\left.+2c_{12}^3c_{13}^2c_{23}s_{12}s_{13}s_{23}\cos(\phi_{12}+\phi_{23}-\phi_{13})-2c_{12}c_{23}s_{13}s_{23}c_{13}^2s_{12}^2\cos(\phi_{12}+\phi_{23}-\phi_{13})\right] , \\
\label{eq28}C_3=&N^2{e}^{-2\gamma(\alpha-\alpha_0)^2} \left[c_{12}^4c_{13}^2c_{23}^2s_{13}^2 + 2c_{12}^2c_{13}^2s_{12}^2s_{23}^2+c_{13}^2c_{23}^2s_{12}^4s_{13}^2 + c_{13}^2c_{23}^2s_{13}^2  \right.\nonumber\\
&\left.+2c_{12}c_{13}^2c_{23}s_{12}s_{13}s_{23}(s_{12}^2-c_{12}^2)\cos(\phi_{12}+\phi_{23}-\phi_{13})\right] .
\end{align}
\end{widetext}

Another important quantity we focus on is the $CP$ violation of neutrinos which plays an important role in the study of weak interactions. The transition amplitude of anti-neutrino, i.e., $\bar{\nu}_{e}\rightarrow\bar{\nu}_{\mu}$, is obtained by inversing the $CP$ violation angle $\phi_{ij}\rightarrow-\phi_{ij}$. From Eq. (\ref{eq25}) we can get the ratio of $CP$ violation of the muon neutrino

\begin{align}
&\rho_{\mu}(\alpha,t)-\rho_{\bar{\mu}}(\alpha,t) = 4N^2c_{12}c_{13}^2c_{23}s_{12}s_{13}s_{23}{e}^{-2\gamma(\alpha-\alpha_0)^2} \nonumber \\
&\sin(\phi_{12}+\phi_{23}-\phi_{13})\left[\sin(\Omega_{12}t)-\sin(\Omega_{13}t)+\sin(\Omega_{23}t)\right] .
\end{align}
The numerical result is shown in Fig. (\ref{fig3}). One can find that the oscillation period is the same as the period of the density oscillations $\lambda_{21}$ shown in Fig. (\ref{fig1}) and (\ref{fig2}). However the maximum value of $CP$ violation is corresponding to the zero point of the density oscillation, i.e., distance $\approx75$km.

The preceding paragraphs are concerned with the propagation behavior for neutrinos with the single value of the energy. In this case, neutrino quantities fluctuate in the space like simple harmonic oscillators with their amplitudes and periods keep fixed.
Now we concentrate on the total effect of neutrinos with the Gaussian-like energy distribution. In Fig. (\ref{fig4}), we show the survival probability of the electron neutrino $SP(x)=\rho_{e}(x)/\rho_{e}(0)$ as a function of the distance away from the source. The Gaussian integration in Eq. (\ref{eq23}) is difficult to be carried out analytically, therefore it was computed numerically by the Romberg method\cite{NR}. When the propagation distance is short ($x\rightarrow0$), one can see from Eq. (\ref{eq23}) that the variation of the survival probability
\begin{equation}
\frac{d{SP}(x)}{xdx}\propto-\Delta{m}_{12}^2\cos^4\theta_{13}\sin^22\theta_{12}-\Delta{m}_{13}^2\sin^22\theta_{13}
\end{equation}
and the term proportional to $\sin^22\theta_{13}$ plays a leading role in the oscillation because $\Delta{m}_{12}^2\ll\Delta{m}_{13}^2$. In this case, the short range survival probability $SP_s(x)$ could be obtained by the approximation of setting $\theta_{12}=\theta_{23}=0$. On the other hand, when the electron neutrinos oscillate to another flavor over a relatively long distance, the long range probability $SP_l(x)$ could be gained by setting $\theta_{13}=0$ due to the fact that $\sin(\theta_{13})\ll\sin(\theta_{12})$. In the plot, it could be seen that $SP_l(x)$ show good agreement with $SP(x)$ when $x$ is large, which indicates the contribution of $\theta_{13}$ is small in this situation. This may explain why $\theta_{13}$ is more difficult to be measured than $\theta_{12}$ and $\theta_{23}$ in long baseline experiments. On the contrary, it also could be deduced from our results that $\theta_{13}$ is easier to be measured than $\theta_{12}$ and $\theta_{23}$ in short baseline experiments. In the Daya Bay reactor neutrino experiment, information obtained from a series of detectors, some near the reactors and some more than a kilometer farther away, are brought together to produce a non-zero value for $\theta_{13}$ with a significance of 5.2 standard deviations. Our results provide a good explanation to their success.

\begin{figure}[tb]
  \centering
      \includegraphics[width=0.9\columnwidth]{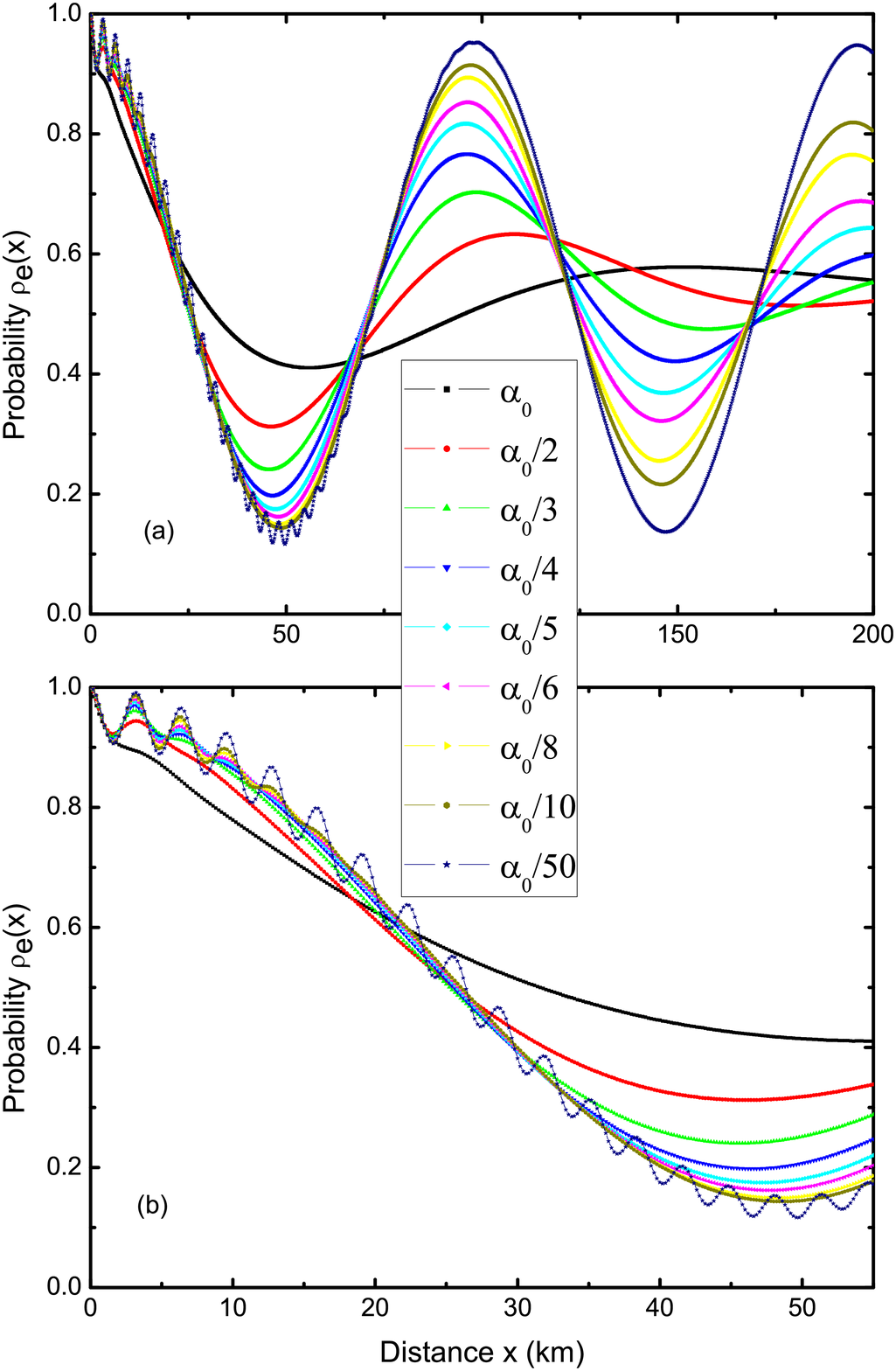}\\
  \caption{The survival probability of the electron neutrino $SP(x)$ as a function of distance for different widths of the Gaussian packet. (a) The long range propagation. The amplitudes decrease with the increasing Gaussian packet width, while the periods remain unaltered. (b) The short range behaviour. The fast oscillations fade out soon for all initial packets. } \label{fig5}
\end{figure}

\begin{figure}[tb]
  \centering
      \includegraphics[width=0.9\columnwidth]{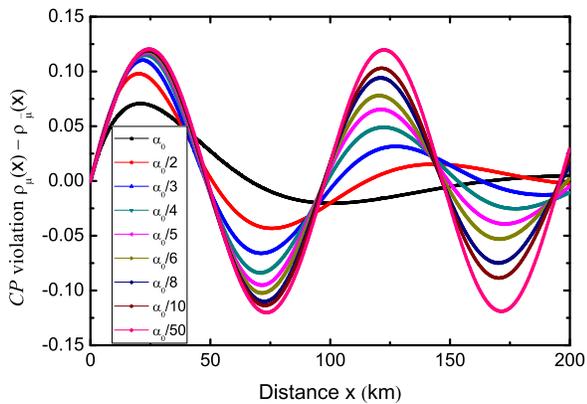}\\
  \caption{The oscillations of the $CP$ violation of the muon neutrinos as a function of distance for different widths of the Gaussian-like packet. } \label{fig6}
\end{figure}

\begin{figure}[tb]
  \centering
      \includegraphics[width=0.9\columnwidth]{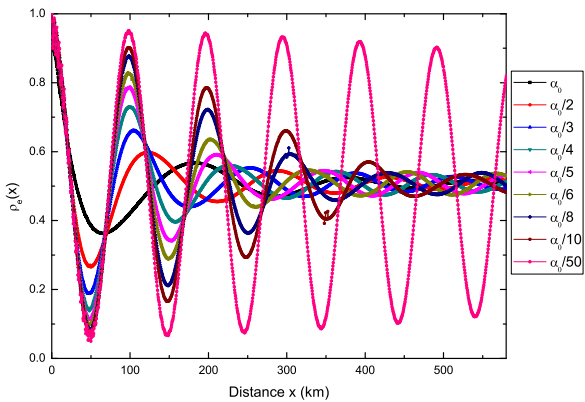}\\
  \caption{The long range behaver of the electron neutrino oscillation for different widths of the Gasuusian packet. } \label{fig7}
\end{figure}

\begin{figure}[tb]
  \centering
      \includegraphics[width=0.9\columnwidth]{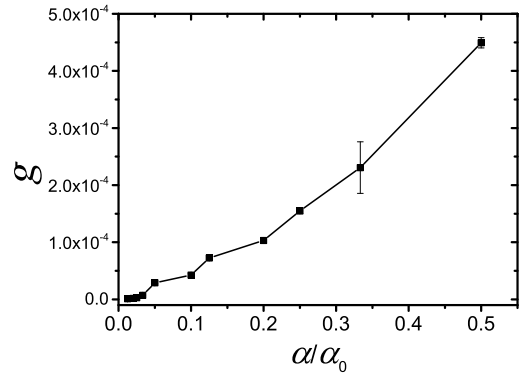}\\
  \caption{The hyper-exponential decay factor $g$ as a function of the widths of the Gaussian-like packet $\alpha/\alpha_0$.} \label{fig8}
\end{figure}

In Fig. (\ref{fig5}) and (\ref{fig6}), we study the survival rates of electron neutrinos and the $CP$ violations of muon neutrinos as a function of distance for neutrino beams with different widths of Gaussian packets. Contrasting to the propagation of the neutrino with single-value energy whose vibration amplitudes are fixed with regard to the distance $x$, the Gaussian-like wave packet disperses in the propagation with its amplitude diminishes and its period increases.
In the meantime, when the source beam has a larger packet width, the fluctuation amplitudes also lowered its value whereas the period increases.
The results indicate that a wide energy distribution of the initial electron neutrino obstructs the detection of neutrino oscillations.
In order to investigate the dependance of neutrino oscillation on the distance $x$, we plot the amplitude of $\rho_e(x)$ versus $x$ for different Gaussian packet widths in Fig. (\ref{fig7}). In order to obtain the quantitative relationship between the oscillation amplitudes and the widths of the Gaussian packet, we fit Fig. (\ref{fig7}) with the decay formula of Gaussian packet:  
\begin{equation} 
\rho_e(x) = \frac{\rho_e(0)}{\sqrt{1+ax^2}}e^{-gx^2/(1+ax^2)},
\end{equation}
where $\rho_e(0)$, $a$, and $g$ are fitting parameters. One can find that the long range behaver of the oscillation amplitude is dominated by the exponential decay factor $\exp{[-gx^2/(a+bx^2)]}$. More importantly, one can find in Fig. (\ref{fig8}) that the hyper-exponential factor $g$ is almost a quadratic fuction of the Gaussian packet width $\alpha/\alpha_0$. These results demonstrate that the signal in short-baseline reactor experiments will be more clear. 

\section{Summary}\label{section4}

In this paper, we investigate the neutrino propagation properties by using the Gaussian-like packet to simulate the energy distribution of the neutrino wave function received by the detector. The time evaluation of the survival probability, the neutrino oscillation, and the $CP$ violation are calculated with the application of the coherent state method. Our results indicate that the mixing angle $\theta_{13}$ is more easier to be measured in short baseline experiments than $\theta_{12}$ and $\theta_{23}$, with which we evidenced the success of the Daya Bay experiment in the determination of $\theta_{13}$. During the propagation, the wave packet diffuses to a wider range and the neutrino oscillation fades. Furthermore, the oscillation amplitude decreases with the increasing Gaussian-like packet width, whereas the oscillation period becomes larger. This suggests that it is more easier to detect neutrino oscillations when utilise a sharper distribution of neutrino energy.




\section{Acknowledgement}

This work is supported by the Special Foundation for Theoretical Physics Research Program of China under the Grant No. 11447167.



\bibliographystyle{apsrev4-1}
\bibliography{ref}

\begin{thebibliography}{20}%
\makeatletter
\providecommand \@ifxundefined [1]{%
 \@ifx{#1\undefined}
}%
\providecommand \@ifnum [1]{%
 \ifnum #1\expandafter \@firstoftwo
 \else \expandafter \@secondoftwo
 \fi
}%
\providecommand \@ifx [1]{%
 \ifx #1\expandafter \@firstoftwo
 \else \expandafter \@secondoftwo
 \fi
}%
\providecommand \natexlab [1]{#1}%
\providecommand \enquote  [1]{``#1''}%
\providecommand \bibnamefont  [1]{#1}%
\providecommand \bibfnamefont [1]{#1}%
\providecommand \citenamefont [1]{#1}%
\providecommand \href@noop [0]{\@secondoftwo}%
\providecommand \href [0]{\begingroup \@sanitize@url \@href}%
\providecommand \@href[1]{\@@startlink{#1}\@@href}%
\providecommand \@@href[1]{\endgroup#1\@@endlink}%
\providecommand \@sanitize@url [0]{\catcode `\\12\catcode `\$12\catcode
  `\&12\catcode `\#12\catcode `\^12\catcode `\_12\catcode `\%12\relax}%
\providecommand \@@startlink[1]{}%
\providecommand \@@endlink[0]{}%
\providecommand \url  [0]{\begingroup\@sanitize@url \@url }%
\providecommand \@url [1]{\endgroup\@href {#1}{\urlprefix }}%
\providecommand \urlprefix  [0]{URL }%
\providecommand \Eprint [0]{\href }%
\providecommand \doibase [0]{http://dx.doi.org/}%
\providecommand \selectlanguage [0]{\@gobble}%
\providecommand \bibinfo  [0]{\@secondoftwo}%
\providecommand \bibfield  [0]{\@secondoftwo}%
\providecommand \translation [1]{[#1]}%
\providecommand \BibitemOpen [0]{}%
\providecommand \bibitemStop [0]{}%
\providecommand \bibitemNoStop [0]{.\EOS\space}%
\providecommand \EOS [0]{\spacefactor3000\relax}%
\providecommand \BibitemShut  [1]{\csname bibitem#1\endcsname}%
\let\auto@bib@innerbib\@empty
\bibitem [{\citenamefont {Olive}(2014)}]{ChinPhysC.38.090001}%
  \BibitemOpen
  \bibfield  {author} {\bibinfo {author} {\bibfnamefont {K.~A.~{\it et al.}.}\
  \bibnamefont {Olive}} (\bibinfo {collaboration} {Particle Data Group}),\
  }\href {\doibase 10.1088/1674-1137/38/9/090001} {\bibfield  {journal}
  {\bibinfo  {journal} {Chin. Phys. C}\ }\textbf {\bibinfo {volume} {38}},\
  \bibinfo {pages} {090001} (\bibinfo {year} {2014})}\BibitemShut {NoStop}%
\bibitem [{\citenamefont {Beringer}\ \emph {et~al.}(2012)\citenamefont
  {Beringer}, \citenamefont {Arguin}, \citenamefont {Barnett},\ and\
  \citenamefont {{\it et al.}}}]{PhysRevD.86.010001}%
  \BibitemOpen
  \bibfield  {author} {\bibinfo {author} {\bibfnamefont {J.}~\bibnamefont
  {Beringer}}, \bibinfo {author} {\bibfnamefont {J.~F.}\ \bibnamefont
  {Arguin}}, \bibinfo {author} {\bibfnamefont {R.~M.}\ \bibnamefont {Barnett}},
  \ and\ \bibinfo {author} {\bibnamefont {{\it et al.}}} (\bibinfo
  {collaboration} {Particle Data Group}),\ }\href {\doibase
  10.1103/PhysRevD.86.010001} {\bibfield  {journal} {\bibinfo  {journal} {Phys.
  Rev. D}\ }\textbf {\bibinfo {volume} {86}},\ \bibinfo {pages} {010001}
  (\bibinfo {year} {2012})}\BibitemShut {NoStop}%
\bibitem [{\citenamefont {Pontecorvo}(1957)}]{Pontecorvo1}%
  \BibitemOpen
  \bibfield  {author} {\bibinfo {author} {\bibfnamefont {B.}~\bibnamefont
  {Pontecorvo}},\ }\href@noop {} {\bibfield  {journal} {\bibinfo  {journal}
  {Sov. Phys. JETP}\ }\textbf {\bibinfo {volume} {17}},\ \bibinfo {pages} {429}
  (\bibinfo {year} {1957})}\BibitemShut {NoStop}%
\bibitem [{\citenamefont {Pontecorvo}(1968)}]{Pontecorvo2}%
  \BibitemOpen
  \bibfield  {author} {\bibinfo {author} {\bibfnamefont {B.}~\bibnamefont
  {Pontecorvo}},\ }\href@noop {} {\bibfield  {journal} {\bibinfo  {journal}
  {Sov. Phys. JETP}\ }\textbf {\bibinfo {volume} {26}},\ \bibinfo {pages} {984}
  (\bibinfo {year} {1968})}\BibitemShut {NoStop}%
\bibitem [{\citenamefont {Maki}\ \emph {et~al.}(1962)\citenamefont {Maki},
  \citenamefont {Nakagawa},\ and\ \citenamefont {Sakata}}]{Maki01111962}%
  \BibitemOpen
  \bibfield  {author} {\bibinfo {author} {\bibfnamefont {Z.}~\bibnamefont
  {Maki}}, \bibinfo {author} {\bibfnamefont {M.}~\bibnamefont {Nakagawa}}, \
  and\ \bibinfo {author} {\bibfnamefont {S.}~\bibnamefont {Sakata}},\ }\href
  {\doibase 10.1143/PTP.28.870} {\bibfield  {journal} {\bibinfo  {journal}
  {Progress of Theoretical Physics}\ }\textbf {\bibinfo {volume} {28}},\
  \bibinfo {pages} {870} (\bibinfo {year} {1962})}\BibitemShut {NoStop}%
\bibitem [{\citenamefont {Gonzalez-Garcia}\ \emph {et~al.}(2014)\citenamefont
  {Gonzalez-Garcia}, \citenamefont {Maltoni},\ and\ \citenamefont
  {Schwetz}}]{gonzalez2014nufit}%
  \BibitemOpen
  \bibfield  {author} {\bibinfo {author} {\bibfnamefont {M.}~\bibnamefont
  {Gonzalez-Garcia}}, \bibinfo {author} {\bibfnamefont {M.}~\bibnamefont
  {Maltoni}}, \ and\ \bibinfo {author} {\bibfnamefont {T.}~\bibnamefont
  {Schwetz}},\ }\href@noop {} {\bibfield  {journal} {\bibinfo  {journal} {URL
  http://www. nu-fit. org}\ } (\bibinfo {year} {2014})}\BibitemShut {NoStop}%
\bibitem [{\citenamefont
  {2.~Gonzalez-Garcia}(2012)}]{JHighEnergyPhys.2012.123}%
  \BibitemOpen
  \bibfield  {author} {\bibinfo {author} {\bibfnamefont {M.~S. J. S.~T.}\
  \bibnamefont {2.~Gonzalez-Garcia}, \bibfnamefont {M.~C.;~Maltoni}},\ }\href
  {\doibase 10.1007/JHEP12(2012)123} {\bibfield  {journal} {\bibinfo  {journal}
  {Journal of High Energy Physics}\ }\textbf {\bibinfo {volume} {2012}},\
  \bibinfo {pages} {1} (\bibinfo {year} {2012})}\BibitemShut {NoStop}%
\bibitem [{\citenamefont {Fogli}\ \emph {et~al.}(2011)\citenamefont {Fogli},
  \citenamefont {Lisi}, \citenamefont {Marrone}, \citenamefont {Palazzo},\ and\
  \citenamefont {Rotunno}}]{PhysRevD.84.053007}%
  \BibitemOpen
  \bibfield  {author} {\bibinfo {author} {\bibfnamefont {G.~L.}\ \bibnamefont
  {Fogli}}, \bibinfo {author} {\bibfnamefont {E.}~\bibnamefont {Lisi}},
  \bibinfo {author} {\bibfnamefont {A.}~\bibnamefont {Marrone}}, \bibinfo
  {author} {\bibfnamefont {A.}~\bibnamefont {Palazzo}}, \ and\ \bibinfo
  {author} {\bibfnamefont {A.~M.}\ \bibnamefont {Rotunno}},\ }\href {\doibase
  10.1103/PhysRevD.84.053007} {\bibfield  {journal} {\bibinfo  {journal} {Phys.
  Rev. D}\ }\textbf {\bibinfo {volume} {84}},\ \bibinfo {pages} {053007}
  (\bibinfo {year} {2011})}\BibitemShut {NoStop}%
\bibitem [{\citenamefont {An}\ \emph {et~al.}(2012)\citenamefont {An},
  \citenamefont {Bai}, \citenamefont {Balantekin},\ and\ \citenamefont {{\it et
  al.}}}]{PhysRevLett.108.171803}%
  \BibitemOpen
  \bibfield  {author} {\bibinfo {author} {\bibfnamefont {F.~P.}\ \bibnamefont
  {An}}, \bibinfo {author} {\bibfnamefont {J.~Z.}\ \bibnamefont {Bai}},
  \bibinfo {author} {\bibfnamefont {A.~B.}\ \bibnamefont {Balantekin}}, \ and\
  \bibinfo {author} {\bibnamefont {{\it et al.}}} (\bibinfo {collaboration}
  {Daya-Bay Collaboration}),\ }\href {\doibase 10.1103/PhysRevLett.108.171803}
  {\bibfield  {journal} {\bibinfo  {journal} {Phys. Rev. Lett.}\ }\textbf
  {\bibinfo {volume} {108}},\ \bibinfo {pages} {171803} (\bibinfo {year}
  {2012})}\BibitemShut {NoStop}%
\bibitem [{\citenamefont {Ahn}\ \emph {et~al.}(2012)\citenamefont {Ahn},
  \citenamefont {Chebotaryov}, \citenamefont {Choi}, \citenamefont {Choi},
  \citenamefont {Choi}, \citenamefont {Choi}, \citenamefont {Jang},
  \citenamefont {Jang}, \citenamefont {Jeon}, \citenamefont {Jeong},
  \citenamefont {Joo}, \citenamefont {Kim}, \citenamefont {Kim}, \citenamefont
  {Kim}, \citenamefont {Kim}, \citenamefont {Kim}, \citenamefont {Kim},
  \citenamefont {Kim}, \citenamefont {Kim}, \citenamefont {Kim}, \citenamefont
  {Lee}, \citenamefont {Lee}, \citenamefont {Lim}, \citenamefont {Ma},
  \citenamefont {Pac}, \citenamefont {Park}, \citenamefont {Park},
  \citenamefont {Park}, \citenamefont {Shin}, \citenamefont {Siyeon},
  \citenamefont {Yang}, \citenamefont {Yeo}, \citenamefont {Yi},\ and\
  \citenamefont {Yu}}]{PhysRevLett.108.191802}%
  \BibitemOpen
  \bibfield  {author} {\bibinfo {author} {\bibfnamefont {J.~K.}\ \bibnamefont
  {Ahn}}, \bibinfo {author} {\bibfnamefont {S.}~\bibnamefont {Chebotaryov}},
  \bibinfo {author} {\bibfnamefont {J.~H.}\ \bibnamefont {Choi}}, \bibinfo
  {author} {\bibfnamefont {S.}~\bibnamefont {Choi}}, \bibinfo {author}
  {\bibfnamefont {W.}~\bibnamefont {Choi}}, \bibinfo {author} {\bibfnamefont
  {Y.}~\bibnamefont {Choi}}, \bibinfo {author} {\bibfnamefont {H.~I.}\
  \bibnamefont {Jang}}, \bibinfo {author} {\bibfnamefont {J.~S.}\ \bibnamefont
  {Jang}}, \bibinfo {author} {\bibfnamefont {E.~J.}\ \bibnamefont {Jeon}},
  \bibinfo {author} {\bibfnamefont {I.~S.}\ \bibnamefont {Jeong}}, \bibinfo
  {author} {\bibfnamefont {K.~K.}\ \bibnamefont {Joo}}, \bibinfo {author}
  {\bibfnamefont {B.~R.}\ \bibnamefont {Kim}}, \bibinfo {author} {\bibfnamefont
  {B.~C.}\ \bibnamefont {Kim}}, \bibinfo {author} {\bibfnamefont {H.~S.}\
  \bibnamefont {Kim}}, \bibinfo {author} {\bibfnamefont {J.~Y.}\ \bibnamefont
  {Kim}}, \bibinfo {author} {\bibfnamefont {S.~B.}\ \bibnamefont {Kim}},
  \bibinfo {author} {\bibfnamefont {S.~H.}\ \bibnamefont {Kim}}, \bibinfo
  {author} {\bibfnamefont {S.~Y.}\ \bibnamefont {Kim}}, \bibinfo {author}
  {\bibfnamefont {W.}~\bibnamefont {Kim}}, \bibinfo {author} {\bibfnamefont
  {Y.~D.}\ \bibnamefont {Kim}}, \bibinfo {author} {\bibfnamefont
  {J.}~\bibnamefont {Lee}}, \bibinfo {author} {\bibfnamefont {J.~K.}\
  \bibnamefont {Lee}}, \bibinfo {author} {\bibfnamefont {I.~T.}\ \bibnamefont
  {Lim}}, \bibinfo {author} {\bibfnamefont {K.~J.}\ \bibnamefont {Ma}},
  \bibinfo {author} {\bibfnamefont {M.~Y.}\ \bibnamefont {Pac}}, \bibinfo
  {author} {\bibfnamefont {I.~G.}\ \bibnamefont {Park}}, \bibinfo {author}
  {\bibfnamefont {J.~S.}\ \bibnamefont {Park}}, \bibinfo {author}
  {\bibfnamefont {K.~S.}\ \bibnamefont {Park}}, \bibinfo {author}
  {\bibfnamefont {J.~W.}\ \bibnamefont {Shin}}, \bibinfo {author}
  {\bibfnamefont {K.}~\bibnamefont {Siyeon}}, \bibinfo {author} {\bibfnamefont
  {B.~S.}\ \bibnamefont {Yang}}, \bibinfo {author} {\bibfnamefont {I.~S.}\
  \bibnamefont {Yeo}}, \bibinfo {author} {\bibfnamefont {S.~H.}\ \bibnamefont
  {Yi}}, \ and\ \bibinfo {author} {\bibfnamefont {I.}~\bibnamefont {Yu}}
  (\bibinfo {collaboration} {RENO Collaboration}),\ }\href {\doibase
  10.1103/PhysRevLett.108.191802} {\bibfield  {journal} {\bibinfo  {journal}
  {Phys. Rev. Lett.}\ }\textbf {\bibinfo {volume} {108}},\ \bibinfo {pages}
  {191802} (\bibinfo {year} {2012})}\BibitemShut {NoStop}%
\bibitem [{\citenamefont {Bernardini}\ and\ \citenamefont
  {Leo}(2004)}]{PhysRevD.70.053010}%
  \BibitemOpen
  \bibfield  {author} {\bibinfo {author} {\bibfnamefont {A.~E.}\ \bibnamefont
  {Bernardini}}\ and\ \bibinfo {author} {\bibfnamefont {S.~D.}\ \bibnamefont
  {Leo}},\ }\href {\doibase 10.1103/PhysRevD.70.053010} {\bibfield  {journal}
  {\bibinfo  {journal} {Phys. Rev. D}\ }\textbf {\bibinfo {volume} {70}},\
  \bibinfo {pages} {053010} (\bibinfo {year} {2004})}\BibitemShut {NoStop}%
\bibitem [{\citenamefont {Nauenberg}(1999)}]{Nauenberg199923}%
  \BibitemOpen
  \bibfield  {author} {\bibinfo {author} {\bibfnamefont {M.}~\bibnamefont
  {Nauenberg}},\ }\href {\doibase
  http://dx.doi.org/10.1016/S0370-2693(98)01556-1} {\bibfield  {journal}
  {\bibinfo  {journal} {Physics Letters B}\ }\textbf {\bibinfo {volume}
  {447}},\ \bibinfo {pages} {23 } (\bibinfo {year} {1999})}\BibitemShut
  {NoStop}%
\bibitem [{\citenamefont {Kayser}(1981)}]{PhysRevD.24.110}%
  \BibitemOpen
  \bibfield  {author} {\bibinfo {author} {\bibfnamefont {B.}~\bibnamefont
  {Kayser}},\ }\href {\doibase 10.1103/PhysRevD.24.110} {\bibfield  {journal}
  {\bibinfo  {journal} {Phys. Rev. D}\ }\textbf {\bibinfo {volume} {24}},\
  \bibinfo {pages} {110} (\bibinfo {year} {1981})}\BibitemShut {NoStop}%
\bibitem [{\citenamefont {Fuji}\ \emph {et~al.}(2006)\citenamefont {Fuji},
  \citenamefont {Matsuura}, \citenamefont {Shibuya},\ and\ \citenamefont
  {S.Y.Tsai}}]{Fuju2006}%
  \BibitemOpen
  \bibfield  {author} {\bibinfo {author} {\bibfnamefont {C.}~\bibnamefont
  {Fuji}}, \bibinfo {author} {\bibfnamefont {Y.}~\bibnamefont {Matsuura}},
  \bibinfo {author} {\bibfnamefont {T.}~\bibnamefont {Shibuya}}, \ and\
  \bibinfo {author} {\bibnamefont {S.Y.Tsai}},\ }\href@noop {} {\enquote
  {\bibinfo {title} {A wave-packet view of neutrino oscillation and pion
  decay},}\ } (\bibinfo {year} {2006}),\ \Eprint
  {http://arxiv.org/abs/hep-ph/0612300} {hep-ph/0612300} \BibitemShut {NoStop}%
\bibitem [{\citenamefont {Kayser}\ and\ \citenamefont
  {Kopp}(2010)}]{Kayser2010}%
  \BibitemOpen
  \bibfield  {author} {\bibinfo {author} {\bibfnamefont {B.}~\bibnamefont
  {Kayser}}\ and\ \bibinfo {author} {\bibfnamefont {J.}~\bibnamefont {Kopp}},\
  }\href@noop {} {\enquote {\bibinfo {title} {Testing the wave packet approach
  to neutrino oscillations in future experiments},}\ } (\bibinfo {year}
  {2010}),\ \Eprint {http://arxiv.org/abs/1005.4081} {1005.4081} \BibitemShut
  {NoStop}%
\bibitem [{\citenamefont {Aharmim}(2008)}]{PhysRevL.101.111301}%
  \BibitemOpen
  \bibfield  {author} {\bibinfo {author} {\bibfnamefont {B.~{\it et al.}.}\
  \bibnamefont {Aharmim}} (\bibinfo {collaboration} {SNO Collaboration}),\
  }\href {\doibase 10.1103/PhysRevLett.101.111301} {\bibfield  {journal}
  {\bibinfo  {journal} {Phys. Rev. Lett.}\ }\textbf {\bibinfo {volume} {101}},\
  \bibinfo {pages} {111301} (\bibinfo {year} {2008})}\BibitemShut {NoStop}%
\bibitem [{\citenamefont {Adamson}(2011)}]{PhysRevL.106.181801}%
  \BibitemOpen
  \bibfield  {author} {\bibinfo {author} {\bibfnamefont {K.~{\it et al.}.}\
  \bibnamefont {Adamson}} (\bibinfo {collaboration} {MINOS Collaboration}),\
  }\href {\doibase 10.1103/PhysRevLett.106.181801} {\bibfield  {journal}
  {\bibinfo  {journal} {Phys. Rev. Lett.}\ }\textbf {\bibinfo {volume} {106}},\
  \bibinfo {pages} {181801} (\bibinfo {year} {2011})}\BibitemShut {NoStop}%
\bibitem [{\citenamefont {Abe}(2011)}]{PhysRevL.107.041801}%
  \BibitemOpen
  \bibfield  {author} {\bibinfo {author} {\bibfnamefont {K.~{\it et al.}.}\
  \bibnamefont {Abe}} (\bibinfo {collaboration} {T2K Collaboration}),\ }\href
  {\doibase 10.1103/PhysRevLett.107.041801} {\bibfield  {journal} {\bibinfo
  {journal} {Phys. Rev. Lett.}\ }\textbf {\bibinfo {volume} {107}},\ \bibinfo
  {pages} {041801} (\bibinfo {year} {2011})}\BibitemShut {NoStop}%
\bibitem [{\citenamefont {Harada}(2006)}]{EurophysL.101.111301}%
  \BibitemOpen
  \bibfield  {author} {\bibinfo {author} {\bibfnamefont {J.}~\bibnamefont
  {Harada}},\ }\href {\doibase 10.1209/epl/i2006-10101-2} {\bibfield  {journal}
  {\bibinfo  {journal} {EuroPhys. Lett.}\ }\textbf {\bibinfo {volume} {75}},\
  \bibinfo {pages} {248} (\bibinfo {year} {2006})}\BibitemShut {NoStop}%
\bibitem [{\citenamefont {Willian H.~Press}\ and\ \citenamefont
  {Flannery}(2007)}]{NR}%
  \BibitemOpen
  \bibfield  {author} {\bibinfo {author} {\bibfnamefont {W.~T.~V.}\
  \bibnamefont {Willian H.~Press}, \bibfnamefont {Saul A.~Teukolsky}}\ and\
  \bibinfo {author} {\bibfnamefont {B.~P.}\ \bibnamefont {Flannery}},\
  }\href@noop {} {\emph {\bibinfo {title} {Numerical Recipes}}},\ \bibinfo
  {edition} {3rd}\ ed.\ (\bibinfo  {publisher} {Cambridge University Press},\
  \bibinfo {year} {2007})\BibitemShut {NoStop}%
\end{thebibliography}%

\end{document}